\documentclass[conference]{IEEEtran}
\IEEEoverridecommandlockouts
\usepackage{cite}
\usepackage{amsmath,amssymb,amsfonts}
\usepackage{algorithmic}
\usepackage{graphicx}
\usepackage{textcomp}
\usepackage{xcolor}
\usepackage{float}
\usepackage{url}
\usepackage{hyperref}
\usepackage{booktabs} 
\usepackage{tikz}

\newcommand\copyrighttext{%
  \footnotesize \textcopyright 2025 IEEE. Personal use of this material is permitted.
  Permission from IEEE must be obtained for all other uses, in any current or future
  media, including reprinting/republishing this material for advertising or promotional
  purposes, creating new collective works, for resale or redistribution to servers or
  lists, or reuse of any copyrighted component of this work in other works.}
\newcommand\copyrightnotice{%
\begin{tikzpicture}[remember picture,overlay]
\node[anchor=south,yshift=10pt] at (current page.south) 
  {\fbox{\parbox{\dimexpr\textwidth-\fboxsep-\fboxrule\relax}{\copyrighttext}}};
\end{tikzpicture}%
}

\begin{document}

\title{SBASH: a Framework for Designing and Evaluating RAG vs.\ Prompt-Tuned LLM Honeypots}

\author{\IEEEauthorblockN{Adetayo Adebimpe, Helmut Neukirchen, Thomas Welsh}
\IEEEauthorblockA{Department of Computer Science, University of Iceland, Reykjavik, Iceland\\
\{ama59, helmut, tomwelsh\}@hi.is}
}

\maketitle
\copyrightnotice
\begin{abstract}
Honeypots are decoy systems used for gathering valuable threat intelligence or diverting attackers away from production systems. Maximising attacker engagement is essential to their utility. However research has highlighted that context-awareness, such as the ability to respond to new attack types, systems and attacker agents, is necessary to increase engagement. Large Language Models (LLMs) have been shown as one approach to increase context awareness but suffer from several challenges including accuracy and timeliness of response time, high operational costs and data-protection issues due to cloud deployment. We propose the \emph{System-Based Attention Shell Honeypot} (SBASH) framework which manages data-protection issues through the use of lightweight local LLMs. We investigate the use of Retrieval Augmented Generation (RAG) supported LLMs and non-RAG LLMs for Linux shell commands and evaluate them using several different metrics such as response time differences, realism from human testers, and similarity to a real system calculated with Levenshtein distance, SBert, and BertScore. We show that RAG improves accuracy for untuned models while models that have been tuned via a system prompt that tells the LLM to respond like a Linux system achieve without RAG a similar accuracy as untuned with RAG, while having a slightly lower latency.
\end{abstract}


\section{Introduction}
Acquiring accurate and actionable threat intelligence is crucial to cybersecurity. Honeypots are an  approach to gathering threat intelligence in which intentionally vulnerable systems are used to lure human or automated attackers to understand their attack vectors and even as decoys from real systems~\cite{ilg2023survey}. Honeypots range from \textit{high-interaction}, which are full systems 
but have high complexity and therefore come with the danger of allowing attackers to escape and attack production systems; or \textit{low-interaction} which are more secure as they are emulated systems yet suffer from a lack of realism~\cite{ilg2023survey}. Balancing realism and security of the honeypot to increase attacker engagement while being secure is a principal factor in honeypot system design. Low-interaction honeypots tend to employ static responses which lack context-awareness and the ability to respond dynamically to attacker requests~\cite{sladic2024llm}. This reduces the realism and prevents the collection of threat intelligence for new and emerging attack types. 

Large Language Models (LLMs) have recently been shown to respond effectively to dynamic user input in a wide variety of contexts including cybersecurity~\cite{zhang2025llms}~\cite{nizon2024towards}, text generation~\cite{wu2025survey} and code generation~\cite{jiang2024survey}. Therefore, LLMs have the potential to make honeypots more dynamic due their natural language processing power and knowledge of computer systems, protocols and languages. However, LLMs suffer from several drawbacks, including increased resource usage and cost (depending on the model size model and deployment), and slow response time. Hallucination is another issue, where the LLM presents false information as fact (caused due to lack of information in training data). All these factors can reduce the realism and indicate to an attacker that the system is a honeypot. 

A wide variety of LLM models are available. We aim to use lightweight local LLMs, i.e.\ models with a low number of trained parameters:
Due to their smaller size, Lightweight LLMs can achieve faster speeds and also allow for local deployment (i.e.\ without the need for an external LLM service). The speed increases may introduce more realism, and keeping the operation within a local system assures privacy by not sending sensitive data to a public cloud as it would be the case for LLMs that are offered as a service. However, the smaller LLM size may reduce accuracy in responses and increase hallucinations. One  approach to reduce hallucinations and to increase accuracy in low parameter models is Retrieval Augmented Generation (RAG)~\cite{yu2024evaluation}. RAG leverages retrieval from an external knowledge base and generation of context-aware and accurate responses, therefore enhancing a generative model through external information.  However, so far, RAG is mostly used in the case of document extraction and summarization and has not been extensively studied for honeypots. 

This work is driven by the following research questions:
\begin{itemize}
	
	\item \textbf{RQ1)} How much does a RAG-based LLM honeypot approach improve accuracy compared to a non-RAG LLM honeypot approach?
	
	\item \textbf{RQ2)} How much does a RAG-based LLM honeypot affect response time compared to a non-RAG LLM honeypot?
	
\item \textbf{RQ3)} How realistic is to a human the output of a RAG-based LLM honeypot?

\end{itemize}

The rest of this paper is structured as follows: following this section, related work is reviewed in the next section. Section~\ref{sec:System Based Attention Shell HoneyPot - A Conceptual Framework} introduces our \emph{System-Based Attention Shell Honeypot} (SBASH) framework for creating LLM honeypots for arbitrary shell-based system types using local inference. A proof-of-concept implementation using the SBASH framework and evaluation results are provided in Section~\ref{sec:Results and Analysis}. Section~\ref{sec:Discussion and Analysis} discusses these results and Section~\ref{sec:Conclusion} concludes the paper.

\section{Related Work}\label{sec:Related Work}
\emph{HoneyLLM}~\cite{guan2024honeyllm} is a shell-based honeypot using in-context learning and leverages LLMs to 
provide realistic output to the user and thus provide context awareness. This approached achieved an 
88\% in terms of response accuracy. It used commercial LLMs such as GPT3.5-turbo, GPT-4o, Claude-2, Claude-3-haiku, and Claude-3-opus. These are large parameter models which means that they are not suitable for on-premise deployment, posing challenges to data security and sovereignty for an organisation in a sensitive area such as cybersecurity.

\emph{LLM in the shell}~\cite{sladic2024llm} is an approach that uses prompt engineering techniques (few-shot prompting~\cite{LanguageModelsAreFewShotLearners2020} combined with chain of thought (COT)~\cite{ChainOfThoughtPromptingElicitsReasoningInLargeLanguageModels2022}) to create more accurate responses. The approach was evaluated on 12 human
participants who executed 76 unique commands and were asked to identify if they were false, achieving 
90\% response accuracy. The authors highlight the cost of their approach using the GPT 3.5-Turbo 16k model as \$0.4 per 30 minutes of active use.


 \emph{Limbosh}~\cite{johnson2024modular} is a shell-based honeypot that has as key feature that it is modular, e.g., arbitrary LLMs can be used as long as they support the OpenAI API\footnote{\url{https://platform.openai.com/docs/overview}}  which includes the possibility to use LLMs that run locally. 
 In comparison to \emph{LLM in the shell}, no prompt engineering approaches are used but rather prompt templating. By using rudimentary prompts, complexity is reduced and applicability to different deployment scenarios is enhanced. Also, prompt injection mitigation techniques are used: input delimiting and input/output guarding. The approach is evaluated by four expert participants, three of them were convinced the shell was real. The participant who identified it as a real system noted it suffered from bugs and had crashed. The low response time of the system was also noted. Accuracies were not provided.

\emph{Decoypot}~\cite{sezgin2025decoypot} is a honeypot that mimics web API responses. Even though it is not a shell-based honeypot, it is listed here as related work because it uses RAG. A two stage semantic retrieval process is used to provide more accurate responses and enhanced context awareness to allow for greater gathering of threat intelligence: firstly, by identifying a similar response from the prompt-request pair and secondly by comparing this selected response against the generated one. This enhances the RAG processes by measuring the similarity score of the response against reference responses and returning the most accurate. This approached achieved an average similarity score of 0.9780.

These existing works highlight the following challenges to LLM based honeypots:
\begin{itemize}
    \item \textbf{Inaccurate responses.} Due to the data a model has been trained on, it can either contain or not contain the needed information as requested by the honeypot system, and due to this limitation, the LLM may hallucinate and give incorrect response. For example, the authors of \emph{LLM in the shell} report such a problem due to lack of information about the presence or absence of the  \texttt{.bashrc} file.
    \item \textbf{Lack of state management.} In the context of a shell-based honeypot, LLMs are unable to properly manage states, e.g., if the working directory is changed (\texttt{cd} command), and later, the \texttt{pwd} command is executed, it might not correctly print the current working directory. Even worse, once the LLMs goes beyond its context window, then it forgets what it has been doing before.
    \item \textbf{Delay in responses.} LLMs’ inferencing requires prediction token by token which introduces a delay in response.
    \item \textbf{High computational resources.} Even the lightest LLMs still need computational power for inferencing thus resulting in high cost if commercial LLM services are used. Previous studies have emphasized the high cost of computation when using  public cloud LLM services.
\end{itemize}

To address these challenges and research gaps, we propose to leverage RAG to provide more dynamic responses while also taking advantage of lightweight LLMs that run locally, to enhance speed and to lower operational costs, while at the same time ensuring enhanced data protection, i.e.\ sensitive data does not flow to public cloud LLM service providers. Independent from these technical challenges, it should also to be noted that the related worked evaluated their solutions only by using a small number of human participants as opposed to using a multi-criteria evaluation that includes qualitative human evaluation as well as quantitative factors such a similarity metrics and speed measurements.

\section{System Based Attention Shell Honeypot - A Conceptual Framework}\label{sec:System Based Attention Shell HoneyPot - A Conceptual Framework}

This section proposes the System-Based Attention Shell (SBASH) framework -- a framework to use a \textit{system type parameter} to direct local LLMs to the required system shell type as needed. It aims to improve realism of a honeypot by ensuring the necessary sources of each system are implemented into the honeypot to broaden its application. For example, it can be used to produce honeypots for the various Linux and MacOS shells and the filesystem layouts specific to these systems as well as for the various Windows shells. 
It supports dynamism to enhance realism and to solve the predictability problem found in traditional honeypots. 
This conceptual framework provides a methodological approach to address the challenges posed to traditional and LLM honeypots. 

\begin{figure*}
    \centering
    \includegraphics[scale=0.1]{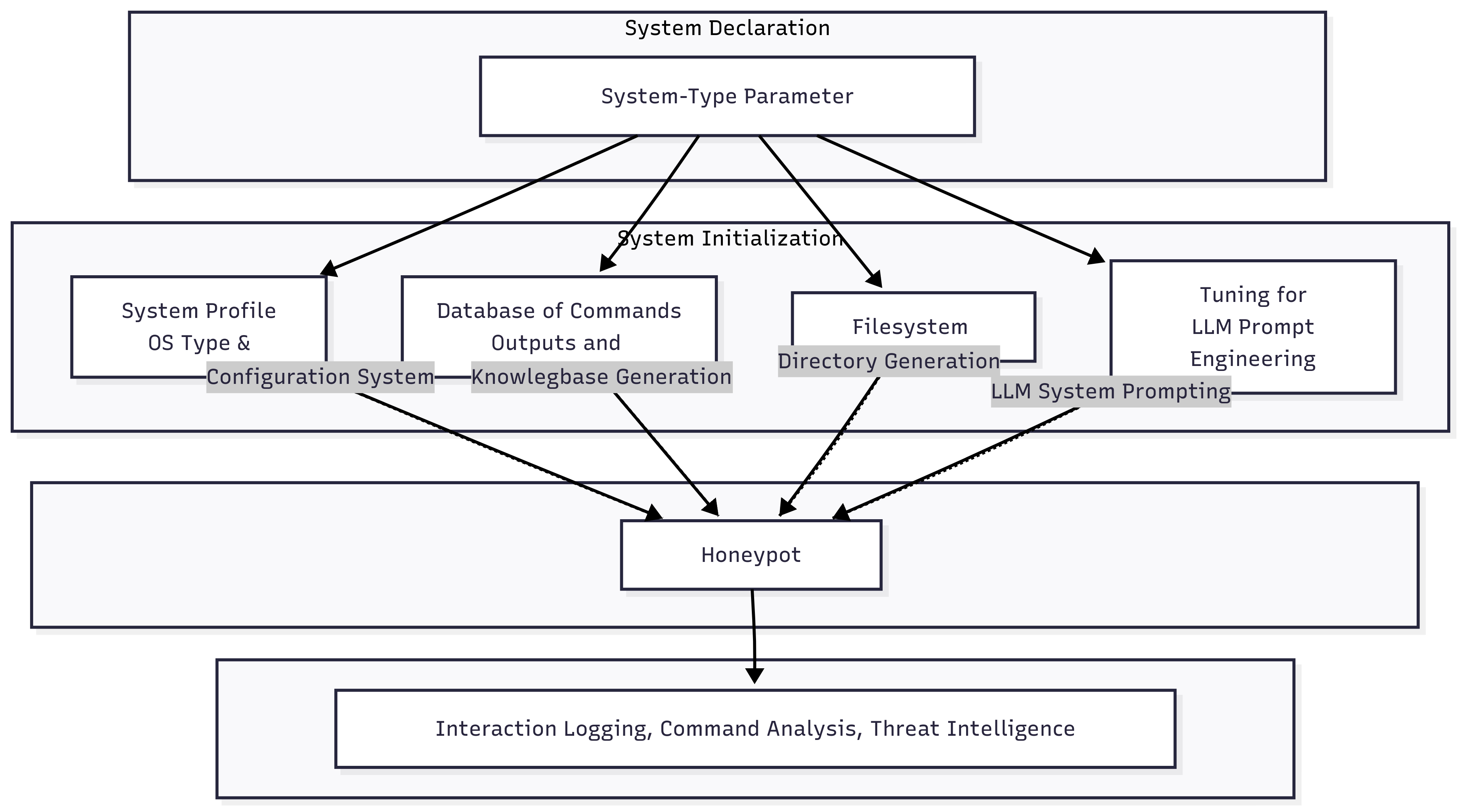}\vspace*{-1ex}
    \caption{SBASH Framework High-level Architecture}\vspace*{-3ex}
    \label{fig:SBASH_framework}
\end{figure*}

As illustrated in \figurename~\ref{fig:SBASH_framework}, the  SBASH framework consists of multiple components:

\begin{itemize}
\item The \textbf{System Declaration} component allows for a centralized parameter control system to allow for easy change of shell types without modifying the main code of the shell-based honeypot. 
\end{itemize}
This centralized parameter is used to initialise the system (e.g.\ hostname, database of commands, filesystem layout, and RAG) by feeding into the four core modules (grey boxes in \figurename~\ref{fig:SBASH_framework}) of the honeypot system: 
\begin{itemize}
\item \textbf{Configuration System:} This component adjusts the system type based on  variables that configure hostname, system users, etc. This allows for consistent data to introduce realism in correlation with the honeypot’s output. 
This component ensures consistency in changes such as the host name across all aspects of the system. 
It handles also the session listener’s configuration which allows for real-time monitoring of an attack.

\item \textbf{Knowledge base generation:} The RAG pipeline requires a source that is referenced during retrieval and generation. Upon the centralized system changes, this module automatically loads the necessary RAG documents specific to the system to be simulated by the honeypot, i.e.\ the shell commands that are supported. This data could include sample executions of each supported shell command or even the manual page that describes that command.

\item \textbf{Directory generation:} File system navigation and modification are necessary for realism of our shell honeypot. A filesystem template is used to populate the filesystem with the necessary default files and directories for the system type. A data structure representing the target file system can be taken from a real instance of that system with additional pseudorandom mutations.
 
\item \textbf{LLM system prompting:} This is one of the core components as it directs the LLM to a better understanding of the request. This allows for realistic output in accordance with the other component configurations. The focus is on the LLM’s preparation and tuning through having a more guided instruct-model that understands the context of the input. This can be achieved via the RAG’s prompt template or by tuning the model’s system prompt (that cannot be overridden by a malicious user prompt)
 itself (see Footnote~\ref{footnote:usedsystemprompt} in Section~\ref{sec:Results and Analysis} for the used system prompt). By doing this, the honeypot system is more context-aware and produces better responses without hallucination. 
\end{itemize}

\noindent
Using these components, the actual honeypot can be deployed and via the configured listener (see Configuration System component), the interactions of the attack with the system can be logged and analysed, supporting threat intelligence.  

A traditional honeypot 
cannot be modified to different needs due to the way it has been shipped with pre-configured files, directories, commands available, etc. In contrast, SBASH has a high adaptability which means we can gather more threat intelligence on different target systems and further improve on realism without breaking the core performance of the system.

\subsection{Processing as imposed by the SBASH framework}

Processing of an attackers shell input is supported by the following stages of the SBASH framework:

\begin{enumerate}
    \item \textbf{Shell Type Parameter Change:} The parameter that determines the shell type (e.g.\ Bash shell) is set once at the beginning to ensure the correct simulation of the needed shell type. 
    \item \textbf{Command Sanitization:} Commands entered by the malicious actor are checked with the list of all default shell commands before sending that command to the RAG for processing -- to avoid prompt injection attacks (e.g.\ sending to the LLM instead of a shell command a question ``Are you an LLM?''). 
    \item \textbf{Command Classification:} Commands are classified into native, AI, and non-existing commands. Native commands get executed without the intervention of AI: usually, these are commands that do not require output, but might, e.g., change file system states, i.e. commands such as  \texttt{cd}, \texttt{mv}, or \texttt{rm}. AI commands get sent to the RAG system for processing. Non-existing commands return a command not found error (with a syntax set by the system type parameter) to the malicious actor. 
    \item \textbf{RAG Processing:} This involves the retrieval and generation, which references the already prepared knowledge base for the usage of the honeypot system. 
    \item \textbf{Threat Intelligence:} During the communication between the user and the system, session and command data (e.g.\ IP used to connect to the system, time and date, command executed, command output) are collected.  
    \item \textbf{Data Evaluation/Analysis:} The stored data are then processed for threat intelligence, but also for the purpose of evaluating the accuracy of the deployed honeypot. 
\end{enumerate}

\section{Implementation and Results}\label{sec:Results and Analysis}


To provide an evaluation of the SBASH system, \emph{HoneywareX}\footnote{\label{github}The source code, implementation details and assets used can be found on GitHub: \url{https://github.com/Cyboghostginx/HoneywareX/tree/main}}, a proof-of-concept, Linux Bash shell honeypot was instantiated using the SBASH framework. This honeypot has been implemented using Python and is utilises standrd Python packages available via PIP, the package installer for Python. Modules like logger, requests, subprocess have been implemented that can be re-used in other honeypot implementations that are based on the SBASH framework. 
We also borrowed the fake filesystem approach from the Cowrie honeypot\footnote{\url{https://github.com/cowrie/cowrie}} that makes file modification and navigation commands as realistic as possible. 

The RAG-enhanced LLM system used in the proof-of-concept implementation consists of four basic parts:

\begin{itemize}
    \item \textbf{Text Embedding:} Text embedding was done using the FASTEmbed model. 
    \item \textbf{Vector Store:} Vector Store used the Llama index vector store core. 
    \item \textbf{Retrieval:} Retrieval used the Llama index retrieval core operation. 
    \item \textbf{LLM Generator:} The RAG generation was done using the 8B parameter Llama model. 
\end{itemize}

The honeypot was evaluated in two different modes (RAG and non-RAG) across three different LLMs ranging from 4B, 8B to 12B parameters, and using LLMs that were 
tuned (by adding a system prompt telling the LLM to behave like a Linux system\footnote{\label{footnote:usedsystemprompt}``\emph{You are a realistic Linux server. Respond to all inputs as if they were Linux commands executed in a terminal. Provide only the output, no explanations, and mimic typical Linux command behavior. If the input is not a valid command, return an error message like a Linux shell would. Don't explain under any condition so you don't expose yourself to the user as AI.}''}) and LLMs without such a system prompt tuning. For the human testing, the 8B parameter tuned Llama 3.1 with RAG was used for the experiment. A predefined set of commands (81 in total) was used for an automated evaluation across all modes and models in respect of shell commands typical for reconnaissance, post exploitation, and exfiltration in cybersecurity. 

The analysis of the system is based on similarity of the honeypot output to a real Linux system Bash shell output, execution time, and human evaluation. Similarity is evaluated using various distance metrics and time difference is calculated using the differences in the response latency. This was done for all LLM models, both with RAG and non-RAG modes. Evaluation by humans is done through manual usage of the honeypot system and subsequent surveys.

\begin{itemize}
\item \textbf{RAG mode:} In this mode, LLMs are pointed to an external knowledge base for guidance and to alleviate hallucination in the case of non-existing data. 
	
\item \textbf{Non-RAG mode}: This mode is a direct inferencing from the LLMs without any external knowledge base. It is only dependent on the data it has been trained on and the system prompts. 
\end{itemize}

\begin{itemize}
	\item \textbf{Gemma 3 (4B) (Tuned and Untuned):} This is a 4B parameter open-source model from Google. 
	
	\item \textbf{Gemma 3 (12B) (Tuned and Untuned):} This is a 12B parameter open-source model from Google. 
	
	\item \textbf{Llama 3.1 (8B) (Tuned and Untuned):} This is an 8B parameter open-source model from Meta. 
\end{itemize}	

\subsection{Accuracy Results}\label{sec:Accuracy Results}

\begin{table*}[!tb]
	\caption{Comparison of RAG vs Non-RAG Accuracy and Tuned Accuracy across different metrics and models.}\vspace{-1.5ex}
	\label{tab:rag_all_accuracy}
	\centering
	\begin{tabular}{llcccc}
        \toprule
		\textbf{Metric} & \textbf{Model} & \textbf{RAG Accuracy (\%)} & \textbf{Non-RAG Accuracy (\%)} & \textbf{RAG Tuned Accuracy (\%)} & \textbf{Non-RAG Tuned Accuracy (\%)} \\
        \midrule
		Levenshtein & Gemma 12B   & 20.7 & 1.4  & 21.9 & 24.6 \\
		Levenshtein & Gemma 4B    & 10.5 & 1.7  & 16.7 & 17.5 \\
		Levenshtein & Llama3.1 8B & 19.0 & 2.9  & 21.3 & 20.0 \\
        \midrule
		SBert       & Gemma 12B   & 47.7 & 35.0 & 49.2 & 56.0 \\
		SBert       & Gemma 4B    & 35.8 & 35.3 & 41.6 & 45.0 \\
		SBert       & Llama3.1 8B & 52.4 & 34.8 & 52.5 & 53.2 \\
        \midrule
		BertScore   & Gemma 12B   & 82.4 & 78.4 & 82.9 & 83.9 \\
		BertScore   & Gemma 4B    & 78.6 & 77.9 & 80.8 & 81.4 \\
		BertScore   & Llama3.1 8B & 83.3 & 78.6 & 83.6 & 83.7 \\
        \bottomrule
	\end{tabular}\vspace*{-4ex}
\end{table*}

 Three similarity  measures were chosen to provide a comparison and understand the efficacy of each one for evaluating accuracy of the responses.:

\begin{itemize}
	\item\textbf{Levenshtein:}~\cite{yujian2007normalized} A distance-based algorithm, it takes insertion, deletion, and substitution of single characters into consideration to calculate the similarity score/percentage. 
	
	\item \textbf{SBert (Sentence-BERT):}~\cite{opitz2022sbert} A semantic embedding paired with cosine similarity to calculate the similarity score of sentences compared to the ground truth. 
	
	\item \textbf{BertScore:}~\cite{zhang2019bertscore} An evaluation metric also derived from the BERT model, but uses an algorithm on both character and word level to compare ground truth and candidate for similarity. This allows for F1 precision recall score (while the SBert score is a dense vector that represents the sentence's semantic meaning). 
\end{itemize}	

Table \ref{tab:rag_all_accuracy} shows the accuracy for the tuned and untuned LLMs with and without RAG using the three similarity measures. It can be seen that the tuned LLMs were more accurate than the non-tuned LLMs. 
It can also be seen that for the untuned LLMs, our RAG approach increases the accuracy significantly. However, for the tuned LLMs, the non-RAG approach provided better overall accuracy than the RAG approach (see discussion in Section~\ref{sec:Discussion and Analysis}); however, this performance difference was marginal.
The larger LLMs perform in general better than smaller LLMs, however the 8B~Llama was often close to the 12B~Gemma and had in some case even a better accuracy.

\subsection{Response Latency}

\begin{figure}[!tb]
    \centering
    \hspace*{-0.95ex}\includegraphics[scale=0.6]{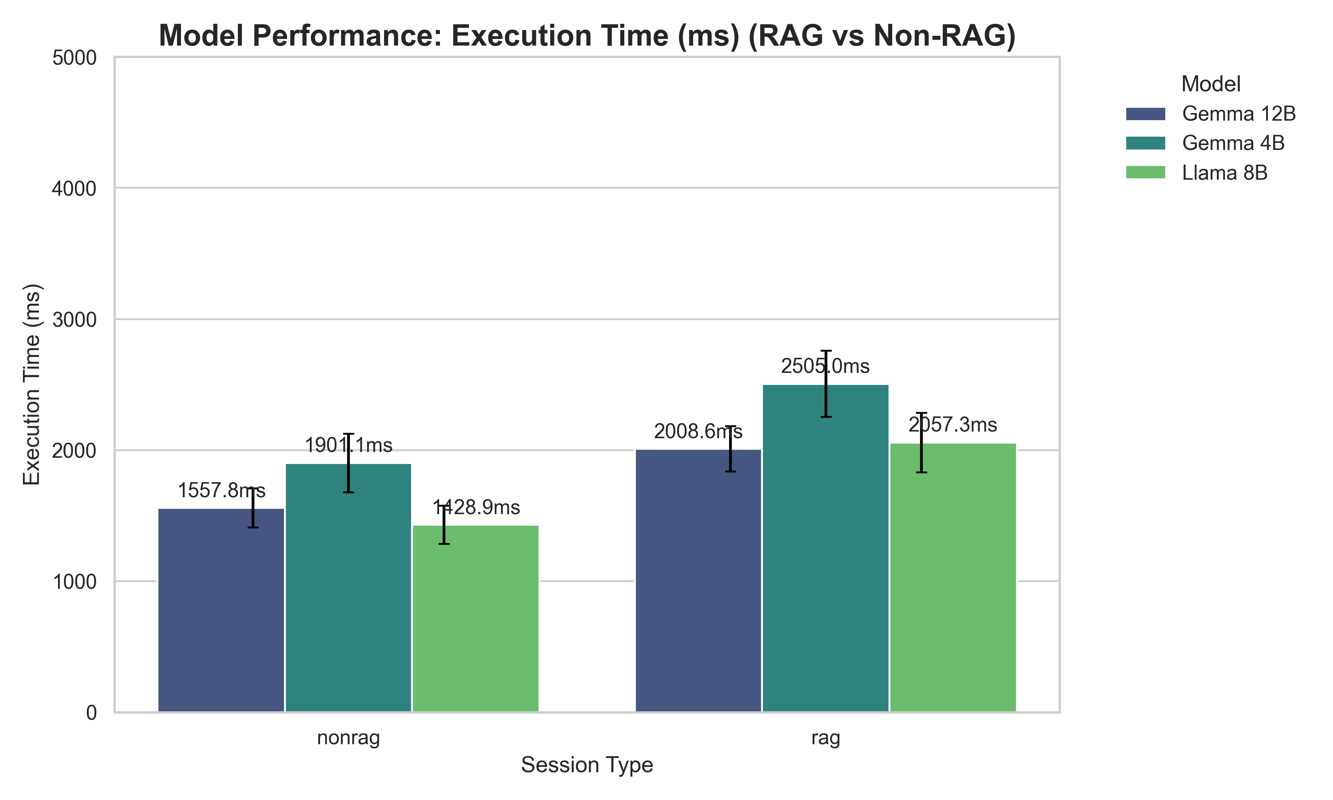}\vspace*{-2.5ex}
    \caption{Model Performance - Execution Time of evaluated models and model sizes with RAG and without RAG} 
    \label{fig:exec_time}
\end{figure}

Latency was measured in milliseconds (ms) using the timer provided by Python. The response latency has been defined as the duration between request and response across all system processes leading to the execution of the commands. 
Response time is one of the important factors that ensures realism in this project.
Due to the different LLMs used, varying response time is expected. Latency was recorded across the three models with and without RAG.
As can be seen in \figurename~\ref{fig:exec_time}, the RAG-based approach had a higher mean latency across all the three models. This can be attributed to the processing time added by the RAG. The model parameter size does not necessarily correlate with increased processing time as the Gemma 12B model had lower latency than Gemma 4B and a marginal difference between Llama 8B: the reason is that the smaller model (Gemma 4B) was less accurate (see Section~\ref{sec:Accuracy Results} and Table~\ref{tab:rag_all_accuracy}) and did therefore tend to create longer output than the larger models which increases execution time.

\subsection{Realism Results}
We evaluated the realism, dependent on human’s judgement (which may differ from the automated similarity metric-based evaluation). Several factors such as speed and accuracy have been evaluated with human judgment. Nine individuals from different field in computer science were selected to give their judgement on the realism of the system.  The participants had varying technical background and were familiar with using a shell. Four people considered themselves as beginner, another four as intermediate, and one person declared having an advanced competency in cybersecurity. The honeypot system was accessed via SSH. 
For realism evaluation, the RAG-enhanced tuned Llama 8B LLM was used. 

\begin{figure}[!tb]
    \centering\vspace*{-2ex}
    \hspace*{-3ex}\includegraphics[scale=0.6]{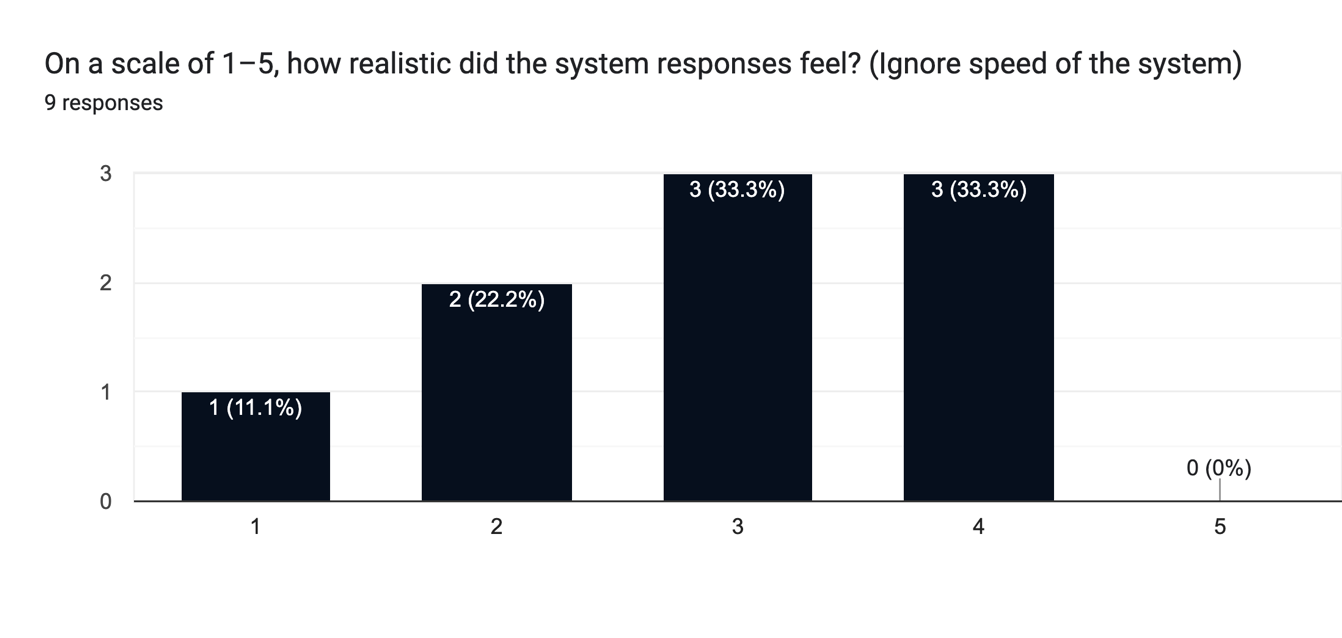}\vspace*{-6ex}
    \caption{Realism rating from human testers: 1=very unrealistic, 5=very realistic}
    \label{fig:realism}
\end{figure}

Each participant was asked to rate the realism of the honeypot on a 5-point Likert scale (with 1=very unrealistic, 5=very realistic), see \figurename~\ref{fig:realism}. After using the system for 5 minutes, they were provided a Google form link to fill out a survey and give their review about the project. 
In addition to answers based on a scale, also open-ended written feedback was collected, and questions about the reasons for their rating were asked. Users could also give feedback on what they think can be used to improve the honeypot system. 

Open-ended answers included, e.g., “\emph{...what mostly makes it unrealistic is just that basic linux CLI commands aren't working. Like arrows to modify a command, tab, up for previous commands. Of course some of these might be disabled in some systems.}” -- which shows that command line editing functionality improvements are needed. 
Another user also pointed out the speed issue which could be a sign that the honeypot was in fact a decoy. 

\subsection{Inference Cost}
Computational cost due to inferencing of the LLMs is an important aspect. In this work, inference time is defined as the time it took for each request to process, from the time of command submission to the full LLM’s output. Any request that requires the inference of the LLMs are considered in this evaluation. Every token output counts towards a computational cost. All inferences were conducted on a single NVIDIA H100 GPU rented on \href{https://modal.com}{modal.com} (for a real, privacy sensitive honeypot, a local GPU would be used). For the human evaluation and the automated evaluation, a total of 3~hours was used: 56~minutes for the human testing, and the remaining time went into the automated evaluation which required using the Python \texttt{subprocess} module to automatically run the commands and save its outputs in a text file. 

The used NVIDIA H100 did cost \$3.95 per hour which means that the total cost of this evaluation were \$11.85 for 3~hours of usage -- extremely low computational cost compared to the previous research pertaining to honeypots requiring LLMs. The mean inference time for non-RAG was shorter compared to RAG. This was expected due to the extra computation that goes into storage and retrieval in RAG. For RAG, an average inference time of 2008.6 ms was recorded for the largest model (Gemma 12B), while 1557.8~ms was recorded in non-RAG for the same model. Gemma 12B, being a 12 billion parameter model, required more computational resources compared to the smallest model (Gemma 4B). This cost of \$0.06 per hour accumulated to \$40 per month. 
 
\section{Discussion}\label{sec:Discussion and Analysis}

\noindent\textbf{Accuracy} (RQ1): 
While RAG excelled for the untuned LLMs, for LLMs instructed to create shell output via a tuned system prompt, the non-RAG approach was slightly better which indicates that the one-shot system prompt tuning technique is better performing than the RAG LLM powered only with a document containing manual/help pages of commands. Further experiments are needed to understand whether a more appropriate RAG document would increase the RAG accuracy. 
\\
\textbf{Response Latency} (RQ2): 
For a realistic shell honeypot, latency should be close to a real system.
As shown in Section~\ref{sec:Results and Analysis}, the average latency on the highest parameter model was a 2~second delay. This is too long to be considered close to a real system.
Moreover, the higher the number of parameter of the LLM, the more resource intensive inferencing would be. 

Speculative decoding \cite{xia2024SpeculativeDecoding} could yield a 2--4 times speed up of the decoding step of an LLM:
 a small, fast model generates a draft sequence of candidate tokens, which are then verified in parallel by the larger, main model in a single forward pass.

We can expect that due to hardware improvements and LLM optimisations, latency will improve in future, so that an LLM-based honeypot's latency becomes closer to the speed of a real, low-resource system, e.g., a home router or IoT system. 

\noindent\textbf{Realism} (RQ3): Some users complain about some interactivity problem such as lack of \texttt{nano} and \texttt{vim} for editing files, arrow keys for commands navigation, and tab for auto-completion. All these shortcomings introduced a limited interactivity in the system which reduces the realism rating for the human’s rating. Despite this, the user still reported average of 4~out~5 for responsiveness.
We have to point out a limitation due to the small number of participants (n=9), but this number is similar to the related works.

In addition to answering the research questions, the following aspects are worth to be discussed:\\
\noindent\textbf{Scalability:} This system has not been put through a test on large scale, like having thousands of requests hit it at once. However we note the ability to scale is feasible but directly constrained by the cost of inference. 

\noindent\textbf{Knowledge Base Limitation:} The RAG document that was used in this system was just the manual pages (\texttt{man} command for all commands in directory \texttt{bin} on an Ubuntu Linux system and their \texttt{-h} option help output); this however, might not be the most effective accurate dataset. More datasets like the input and output of all commands would be more effective to ensure more accurate responses.

\section{Conclusion}\label{sec:Conclusion}

In this work we presented SBASH, a framework to build shell based honeypots for arbitrary systems: just by changing configuration parameters (such as reference outputs of commands and filesystem layout), different operating systems and shells can be simulated. The framework takes advantage of RAG to enhance the accuracy of low parameter LLMs that are used to be able to run them locally to mitigate data protection issues in public clouds. We illustrated that RAG improves accuracy in untuned models at the expense of slightly longer response times. However, system prompt tuning did lead to better accuracy than RAG while having shorter response times.

We observed a sweet spot in model size with respect to response latency and accuracy: too large models are slower, but more accurate, while too small models are less accurate leading to verbose output slowing down response latency. 

To the best of our knowledge, this is the first evaluation of the performance of RAG-based LLMs for shell-based honeypots and the first multi-factor evaluation that takes similarity measures \emph{and} human evaluation into account (in addition to measuring speed). In future work we aim to introduce line editing improvements suggested by the human evaluators and to investigate the influence of an improved RAG document.

\section*{Acknowledgements}
This project has received co-funding from the European Union's Digital Europe Programme under grant agreement no.\ 101127453 National Coordination Centre for Cybersecurity in Iceland and 101127307 Defend Iceland: Nationwide bug bounty platform.

\bibliographystyle{IEEEtran}
\bibliography{refs}

\begin{thebibliography}{10}
\providecommand{\url}[1]{#1}
\csname url@samestyle\endcsname
\providecommand{\newblock}{\relax}
\providecommand{\bibinfo}[2]{#2}
\providecommand{\BIBentrySTDinterwordspacing}{\spaceskip=0pt\relax}
\providecommand{\BIBentryALTinterwordstretchfactor}{4}
\providecommand{\BIBentryALTinterwordspacing}{\spaceskip=\fontdimen2\font plus
\BIBentryALTinterwordstretchfactor\fontdimen3\font minus
  \fontdimen4\font\relax}
\providecommand{\BIBforeignlanguage}[2]{{%
\expandafter\ifx\csname l@#1\endcsname\relax
\typeout{** WARNING: IEEEtran.bst: No hyphenation pattern has been}%
\typeout{** loaded for the language `#1'. Using the pattern for}%
\typeout{** the default language instead.}%
\else
\language=\csname l@#1\endcsname
\fi
#2}}
\providecommand{\BIBdecl}{\relax}
\BIBdecl

\bibitem{ilg2023survey}
N.~Ilg, P.~Duplys, D.~Sisejkovic, and M.~Menth, ``A survey of contemporary
  open-source honeypots, frameworks, and tools,'' \emph{J. Netw. Comput.
  Appl.}, vol. 220, 2023.

\bibitem{sladic2024llm}
M.~Sladi{\'c}, V.~Valeros, C.~Catania, and S.~Garcia, ``{LLM} in the shell:
  Generative honeypots,'' in \emph{Europ. Symp. Secur. Priv. Workshops
  (EuroS\&P)}.\hskip 1em plus 0.5em minus 0.4em\relax IEEE, 2024.

\bibitem{zhang2025llms}
J.~Zhang \emph{et~al.}, ``When {LLMs} meet cybersecurity: A systematic
  literature review,'' \emph{Cybersecurity}, vol.~8, no.~1, 2025.

\bibitem{nizon2024towards}
M.~Nizon-Deladoeuille, B.~Stef{\'a}nsson, H.~Neukirchen, and T.~Welsh,
  ``Towards supporting penetration testing education with large language
  models: an evaluation and comparison,'' in \emph{Int. Conf. Soc. Netw.
  Analysis Manag. Secur. (SNAMS)}.\hskip 1em plus 0.5em minus 0.4em\relax IEEE,
  2024.

\bibitem{wu2025survey}
J.~Wu \emph{et~al.}, ``A survey on {LLM}-generated text detection: Necessity,
  methods, and future directions,'' \emph{Comput. Linguist.}, vol.~51, no.~1,
  2025.

\bibitem{jiang2024survey}
J.~Jiang, F.~Wang, J.~Shen, S.~Kim, and S.~Kim, ``A survey on large language
  models for code generation,'' \emph{ACM Trans. Softw. Eng. Methodol.}, Jul.
  2025.

\bibitem{yu2024evaluation}
H.~Yu \emph{et~al.}, ``Evaluation of retrieval-augmented generation: A
  survey,'' in \emph{BigData 2024}.\hskip 1em plus 0.5em minus 0.4em\relax
  Springer, 2025.

\bibitem{guan2024honeyllm}
C.~Guan, G.~Cao, and S.~Zhu, ``{HoneyLLM}: Enabling shell honeypots with large
  language models,'' in \emph{Conf. Commun. Netw. Secur. (CNS)}.\hskip 1em plus
  0.5em minus 0.4em\relax IEEE, 2024.

\bibitem{LanguageModelsAreFewShotLearners2020}
T.~B. Brown \emph{et~al.}, ``Language models are few-shot learners,'' in
  \emph{Proc. Int. Conf. Neural Inf. Proc. Syst. (NIPS)}.\hskip 1em plus 0.5em
  minus 0.4em\relax Curran Assoc., 2020.

\bibitem{ChainOfThoughtPromptingElicitsReasoningInLargeLanguageModels2022}
J.~Wei \emph{et~al.}, ``Chain-of-thought prompting elicits reasoning in large
  language models,'' in \emph{Proc. Int. Conf. Neural Inf. Proc. Syst.
  (NIPS)}.\hskip 1em plus 0.5em minus 0.4em\relax Curran Assoc., 2022.

\bibitem{johnson2024modular}
S.~Johnson, R.~Hassing, J.~Pijpker, and R.~Loves, ``A modular generative
  honeypot shell,'' in \emph{Int. Conf. Cyber Secur. Resil. (CSR)}.\hskip 1em
  plus 0.5em minus 0.4em\relax IEEE, 2024.

\bibitem{sezgin2025decoypot}
A.~Sezgin and A.~Boyac{\i}, ``{DecoyPot}: A large language model-driven web
  {API} honeypot for realistic attacker engagement,'' \emph{Comput. Secur.},
  vol. 154, 2025.

\bibitem{yujian2007normalized}
L.~Yujian and L.~Bo, ``A normalized {Levenshtein} distance metric,'' \emph{IEEE
  Trans. Pattern Anal. Mach. Intell.}, vol.~29, no.~6, 2007.

\bibitem{opitz2022sbert}
J.~Opitz and A.~Frank, ``{SBERT} studies meaning representations: Decomposing
  sentence embeddings into explainable semantic features,'' in \emph{Proc. 2nd
  Conf. Asia-Pac. Chapter Association Comput. Linguist. 12th Int. Jt. Conf.
  Nat. Lang. Proces. (2nd AACL 12th IJCNLP)}.\hskip 1em plus 0.5em minus
  0.4em\relax Assoc. Comput. Linguist. (ACL), 2022.

\bibitem{zhang2019bertscore}
T.~Zhang, V.~Kishore, F.~Wu, K.~Q. Weinberger, and Y.~Artzi, ``{BERTScore}:
  Evaluating text generation with {BERT},'' in \emph{Int. Conf. Learn.
  Represent. (ICLR)}.\hskip 1em plus 0.5em minus 0.4em\relax OpenReview.net,
  2020.

\bibitem{xia2024SpeculativeDecoding}
H.~Xia \emph{et~al.}, ``Unlocking efficiency in large language model inference:
  A comprehensive survey of speculative decoding,'' in \emph{Findings Assoc.
  Comput. Linguist. (ACL)}.\hskip 1em plus 0.5em minus 0.4em\relax Assoc.
  Comput. Linguist. (ACL), 2024.

\end{thebibliography}
\end{document}